\documentstyle[prb,aps]{revtex}

\begin{document}
\draft

\twocolumn[\hsize\textwidth\columnwidth\hsize\csname @twocolumnfalse\endcsname

\title
{
\bf A simple classical mapping of the spin-polarized quantum electron gas:
distribution functions and local-field  corrections.
}
\author
{
\bf M.W.C. Dharma-wardana\cite{byline1} and F. Perrot
}
\address
{
National Research Council, Ottawa,Canada. K1A 0R6\\
Centre d'Etudes de Bruy\`eres le Ch\^atel, P. O. Box 12,
 91689  Bruy\`eres le Ch\^atel, France
}
\date{24 May 1999}
\maketitle
\begin{abstract}
We use the now well known  spin {\it unpolarized}$\,$
exchange-correlation energy $E_{xc}$ of the uniform
electron gas as the basic ``many-body'' input to
determine the  temperature  $T_q$ of
a {\it classical}$\,$
Coulomb fluid having the same correlation energy
 as the quantum system.
It is shown that the spin-polarized pair distribution functions
(SPDFs) of the
classical fluid at $T_q$, obtained using the hyper-netted chain (HNC)
equation are in excellent agreement with those of the
$T$ = 0 quantum fluid obtained by quantum Monte Carlo
(QMC) simulations. These methods are computationally
simple and  easily applied to problems which are
currently outside the scope of QMC.
Results are presented for the SPDFs and the
local-field corrections to the rsponse functions of the
electron fluid at zero and finite temperatures.
\end{abstract}
%
\vskip2pc]
\narrowtext
	The uniform interacting electron gas (UEG) is the
``text-book'' many-body problem of Fermi liquids in
metals, plasmas or doped semiconductors.\cite{textbk}
 It  provides a model
exchange-correlation potential for the density functional theory
of inhomogeneous
electronic systems.\cite{K-S} 
The random-phase-approximation (RPA) to the properties of the UEG
provides an important 
``baseline'' which could be reached by many
techniques like quantum-linked cluster
expansions, Green's functions, or linearized equations of motion.
RPA is a reasonable approximation when the ''coupling parameter''
$\Gamma$ = (potential energy)/(kinetic energy) of the Coulomb fluid
is smaller
than unity.  The coupling parameter for the UEG at T=0
becomes identical with the mean sphere radius $r_s$ per electron,
i.e., for the 3-D case,
 $\Gamma$ = $r_s$ = $(3/4\pi \overline{n})^{1/3}$. Here
$\overline{n}$ is the electron number-density per atomic unit of volume.
 
Improving the RPA 
becomes nontrivial if the results
of the theory are expected to satisfy sum rules and provide
physically realistic pair-distribution functions (PDFs), i.e.,
 $g(r)$, of the quantum system.
Diagrammatic methods look for resummations that conserve the
 sum rules, Ward identities etc. This line of research is 
followed in the work of
Hubbard, Vosko, Langreth, Geldart et al.\cite{gt}.
Calculations have been
given by MacDonald et al.\cite {ahm} for the self-energy, while Green
et al., Ichimaru et al.
and Richardson et al.
and many others have studied the response functions.\cite{green}

 The equation of motion (EQM) method,
 when pushed
beyond the RPA requires some ``ansatz'' to decouple or 
``close'' the equations.\cite{toigo}
Singwi et al. (SSTL)
used a physically motivated classical analogy to close the
EQM by introducing the electron-electron
PDF determined self-consistently.
There are also attempts to directly {\it fit}$\,$ the
response functions or the PDF's to a form
constrained to satisfy the sum rules and such
requirements.\cite{italians} In spite of a large
effort, the calculation of the electron-gas $g(r)$
by STLS and other methods  at arbitrary
spin-polarization ($\zeta$) and $T$ is
 difficult. In fact,
STLS gives negative PDFs for sufficiently large $r_s$ in the metallic
range and does not satisfy the compressibility sum rule.
Even the attempts to directly fit the PDFs
{\it do not}$\,$ guarantee  positive-definite a $g(r)$ for important
regimes of $r_s$ and $T$.\cite{italians}

	Another approach  beyond RPA is to use a correlated 
wavefunction $\psi=FD$ where $F$ is a correlation factor and $D$
is a Slater determinant. The best  $F$ is
determined  variationally.
This approach leads to the Jackson-Feenberg energy
functional which can be examined in several ways.
\cite{feenberg,lantto,KP} The QMC
techniques also use such a $\psi$ and lead, e.g., to the
variational Monte Carlo (VMC) method.\cite{ceperley81,ortiz,mcs}
Thus Ceperley and
Alder provided $E_{xc}(r_s)$, now available in several 
parametrized forms.\cite{vwn,p-z}
The correlation energy given by various QMC 
methods agree with one another
 to about the same extent as with,
say, the results from the Fermi-Hyper-Netted-Chain(FHNC)
 method.\cite{lantto}. Thus, while it is easy to get good
corrlation energies, it is quite difficult to get other propertis
like the PDFS and local-field corrections (LFC) at
arbitrary $\zeta$, $T$ and in the metallic range of $r_s$.

In  this letter we present a
computationally
simple 
method for calculating the PDFs and other derived
quantities (e.g, response functions) of the UEG, given the 
unpolarized $E_c$ from {\it whatever} source.
 Since PDFs of classical fluids are easily
obtainable via the Hyper-Netted-Chain (HNC) procedure,\cite{hncref}
we  model the electron gas as a {\it classical}$\,$ fluid containing
two species (up, down spins).
We ask for the temperature $T_q$
at which the excess free energy of a  Coulomb fluid,
obeying the {\em classical} HNC-integral equation,
becomes equal to the correlation energy $E_c$ of the UEG at the same
density and at $T$=0. The suffix $q$ in $T_q$ signifies that this temperature
parametrizes the quantum many-body interactions in the UEG.
Using the 
 temperature $T_q$  at each $r_s$  we obtain
 $g_{ij}(r)$, spin-polarized
correlation energies $E_c(r_s,\zeta)$, local-field corrections etc.
While many interesting applications are possible, here
we treat the 3-D spin-polarized UEG at zero and finite $T$.

The physical motivation for our approach comes from density
functional theory (DFT) where the interacting electron gas is replaced
by a noninteracting gas of Kohn-Sham (KS) particles whose wavefunction is
a simple determinant. Here the philosophy is very different
to the Feenberg approach which uses a correlated determinent for 
fermions.\cite{lantto,KP}
In DFT the many-body potential is replaced by a single-particle
KS potential.
Since the natural energy parameter of the clasical ensemble is
the temperature, we look for a temperature
representation of the interactions.

Consider a fluid of mean density $\overline{n}$ containing
two spin species
with concentrations
 $x_i$ = $\overline{n}_i/\overline{n}$. 
In the following we deal with the physical temperature $T$ of
the UEG, while the temperature of the classical fluid $T_{cf}$  is $1/\beta$.
Since the leading dependence of the energy on temperature is quadratic,
we assume that $T_{cf}$= ${\surd{(T^2+T_q^2)}}$. This is
clearly valid for $T=0$ and for very high $T$. This assumption will not be
probed more deeply in this letter where the main effort is for T=0.
Our objective is to determine $T_q$, the temperature
equivalent of the quantum correlations in the UEG at T=0.

The pair-distribution functions for a classical
fluid at an inverse temperature $\beta$ can be written as
\begin{equation}
g_{ij}(r)=exp[\beta \phi_{ij}(r)
-h_{ij}(r)-c_{ij}(r) + B_{ij}(r)]
\label{hnc}
\end{equation}
Here $\phi_{ij}(r)$ is the pair-interaction potential between
species $i,j$. For two electrons this is
just the Coulomb potential $V_{cou}(r)$ if there is no exchange interaction.
If the spins are parallel, the Pauli
principle prevents them from occupying the same spatial orbital.
Following the earlier work, notably by Lado,\cite{lado}
 we introduce a
``Pauli potential'', ${\cal P}(r)$.
Thus:
\begin{equation}
\label{pots}
\phi_{ij}(r)={\cal P}(r)\delta_{ij}+V_{cou}(r)
\end{equation}
The Pauli potential ${\cal P}(r)$ will be discussed 
with the PDFs of the non-interacting UGE, i.e., 
$g^0_{ij}(r)$.
The function $h(r)$ = $g(r)$ -1; it is related to the
structure factor $S(k)$ by a Fourier transform.
The  $c(r)$ is the ``direct correlation function (DCR)''
of the Ornstein-Zernike (OZ)
 equations.
\begin{equation}
\label{oz1}
 h_{ij}(r)  = c_{ij}(r)+
\Sigma_ s\overline{n}_s\int d{\bf r}'h_{i,s}
(|{\bf r}-{\bf r}'|)c_{s,j}({\bf r})
\end{equation}
The $B_{ij}(r)$ term in  Eq.~\ref{hnc} is
the ``bridge'' term arising from certain cluster interactions.
If this is neglected
Eqs.~\ref{hnc}-~\ref{oz1} form a closed set providing the
HNC approximation to the PDF of a classical fluid. 
Various studies have 
clarified the role of $B(r)$ and its treatment
via ``reference'' HNC equations.\cite{rosen}
$B(r)$ is important when the
coupling constant $\Gamma$ exceeds, say, 20.  The range of $\Gamma$ relevant
to this work (e.g., $\Gamma \sim 4.5$ even for $r_s$ = 10 )
is such that the HNC-approximation holds. The HNC-approximation suffers
from a compressibility inconsistency (CI), i.e., the excess compressibility
calculated from the small-$k$ limit of the short-ranged part of c(k)
does not agree with that obtained from
the excess free energy. This CI can be corrected by choosing
$B_{ij}(r)$ suitably.

Consider the non-interacting system at temperature $T$.
The parallel-spin PDF,
i.e, $g_{ii}^0(r,T)$, will be denoted by
$g_T^0(r)$ for simplicity, since $g_{ij}^0(r,T)$, i $\ne$ j is unity.
Consider the paramagnetic case $x_i$ = 0.5.
Denoting $({\bf r}_1-{\bf r}_2)$ by ${\bf r}$, it is easy to show that
\begin{equation}
\label{gzeroT}
g_T^0({\bf r})
=\frac{2}{N^2}\Sigma_{{\bf k}_1,{\bf k}_2}n(k_1)n(k_2)
[1-e^{i({\bf k}_1-{\bf k}_2){\bf \cdot}{\bf r}}]
\end{equation}
Here $n(k)$ is the Fermi occupation number at 
the temperature $T$. Eq.~\ref{gzeroT} reduces to:
\begin{eqnarray}
\label{gzeroTT}
g^0_T(r)&=&1-F_T^2(r)\\
F_T(r)&=&(6\pi^2/k_F^3)\int n(k)\frac{sin(kr)}{r}\frac{kdk}{2\pi^2}.
\end{eqnarray}
The Fermi momentum is denoted by  $k_F$.
Thus $g^0_T(r)$ can be obtained from the Fourier transform of the
 Fermi function. 
Then $c^0(r)$  can be evaluated from
$g^0_T(r)$ using the OZ relations.
The T=0 case can be evaluated analytically.
\cite{kittel}

Assuming that $g^0_{ii}(r)$ can be modeled by an HNC fluid with
the pair interaction $\beta{\cal{P}}(r)$, dropping the indices, we have:
\begin{equation}
g^0(r)=exp[-\beta{\cal{P}}(r)+h^0(r)-c^0(r)]
\end{equation}
The k-space direct correlation  function $c^0(k)$ at T=0
decays as $4k_F/3k$ for small $k$,
 showing that the r-space form $c^0(r)$ is
long ranged. The ``Pauli potential'' ${\cal{P}}(r)$ is given by
\begin{equation}
\label{paudef}
\beta{\cal{P}}(r)=-log[g^0(r)]+h^0(r)-c^0(r)
\end{equation}
We can  determine only the
product $\beta{\cal{P}}(r)$. The classical fluid ``temperature'' $1/\beta$
is still undefined and is {\em not} the thermodynamic
temperature $T$. The Pauli potential
is a universal function of $rk_F$ at $T$=0.
Plots of $\beta{\cal{P}}(r)$ and related functions are given in 
Fig. 1. It is long ranged and mimics 
the exclusion effects of Fermi statistics. 
As $T$ is increased, the Pauli potential 
restricts to about a thermal wavelength
and becomes more hard-sphere like.

	The next step is to use the full pair-potential
$\phi_{ij}(r)$,  Eq.~\ref{pots}, 
and solve the coupled HNC and OZ equations for the binary 
(up and down spins) {\it interacting}$\,$ fluid.
For the paramagnetic case,
$\overline{n}_i$ = $\overline{n}/2$, we have:
\begin{eqnarray}
g_{ij}(r)&=&e^{-\beta{\cal{P}}(r)\delta_{ij}+V_{cou}(r)+h_{ij}(r)-c_{ij}(r)}\\
h_{ij}(q)&=&\stackrel{FT}{\rightarrow}h_{ij}(r)\\
h_{11}(q)&=&c_{11}(q)+(\overline{n}/2)[c_{11}(q)h_{11}(q)+c_{12}(q)h_{21}(q)]\\
h_{12}(q)&=&c_{12}(q)+(\overline{n}/2)[c_{11}(q)h_{12}(q)+c_{12}(q)h_{22}(q)]
\end{eqnarray}
The Coulomb potential $V_{cou}(r)$
needs some
discussion. For two point-charge electrons this is  $1/r$.
However, depending
on the temperature $T$, an electron is 
localized to within a thermal wavelength. Thus, following 
Morita, and Minoo et al.,\cite{minoo} we use a ``diffraction
corrected'' form:
\begin{equation}
\label{potd}
V_{cou}(r)=(1/r)[1-e^{-r/\lambda_{th}}];\;
\lambda_{th}=(2\pi\overline{m}T_{cf}).
\end{equation}
Here $\overline{m}$ is the reduced mass of the electron pair, i,e,
 $m^*(r_s)/2$ a.u., where $m^*(r_s)$, the effective electron mass.  It
 is weakly $r_s$ dependent, e.g, $\sim$0.96 for $r_s$ = 1. In this
 work we take $m^*$=1. The ``diffraction correction'' ensures the
 correct behaviour of $g_{12}(r\sim 0)$ for all $r_s$.

In solving the above equations for a given $r_s$ and at $T$=0,
we have $T_{cf}$=$T_q$. A 
trial $T_q$ is adjust to obtain an $E_c(T_q)$
equal to the known {\it paramagnetic}$\,$ $E_{c}(r_s)$ at each $r_s$.
\begin{equation}
E_c(T_q)=\int_0^1d\lambda\int\frac{1}{r}[h_{11}(r,\lambda)+h_{12}(r,\lambda)]
4\pi r^2dr
\end{equation} 
The resulting
 ``quantum''
 temperatures $T_q$  could be fitted to the form:
\begin{equation}
T_q/E_F=1.0/(a+b{\surd{r_s}}+cr_s)
\label{tfit}
\end{equation}
 The QMC results for $E_{c}$ obtained from different
 QM methods differ, e.g., by
$\sim$6\% at $r_s$=1.
We used the most-recent Ortiz-Ballone
$E_c$ data for the paramagnetic UEG
from VMC and DMC.\cite{ortiz} The small difference in $E_c$ in these
two simulations lead to slightly different fits. The
fit coefficients are, for DMC, a=1.594, b= -0.3160 and c=0.0240,
while for VMC a=1.3251, b= -0.1779
and c=0.0.
Eight values of $r_s$, viz.,
 $r_s$ = $1-6$, 8, and 10 were used in the fit to $T_q$. At $r_s$ =1
and 10,
$T_q/E_F$ goes from 0.768 to 1.198.
 As $r_s\to 0$, $g(r)$ tends to $g^0(r)$.
The  UEG as $r_s\to 0$ goes to a high-density fluid interacting
via the Pauli potential. It
is beyond the HNC-approximation.

	For any given $r_s$, knowing the $T_q$ from the spin-unpolarized
case, we can  obtain
$g_{ij}(r)$ and $E_{xc}(r_s,\zeta, T)$,\cite{p-w} at {\it arbitrary}$\,$,
 spin-polarization $\zeta$
by solving
the coupled HNC equations. Unlike in many of the standard theories of
electron fluids, the PDFs obtained from the HNC-procedure 
are guaranteed to be positive at all $r_s$. In Fig.~2 we show typical
results for $g_{ij}(r)$ and 
compare them with those of QMC-simulations. Our results are
in excellent agreement with the DMC results.
The reported  $g_{12}(r)$  {\em do not} go to
zero for r=0.  The depletion hole for
$g_{12}(r=0)$ from DMC are in agreement with ours 
and are deeper
than those from VMC.\cite{ortiz} The difference in VMC and DMC, even at
$r_s$ =1, Fig.~2(a), is a warning that the even when $E_c$ are in close 
agreement, other properties may have significant errors.
In Fig.~2(c)  we
show the paramagnetic distribution functions at $r_s$ =1 and 10
together with those of the DMC simulations.

At present there is no reliable finite-temperature microscopic
theory to compare with the $g(r,T)$ obtained by our HNC approach.
The theory of Tanaka and Ichimaru,\cite{green} does
better than STLS and is 
comprehensive. However, like in STLS,
the $g(r)$ becomes negative for some values of $r$  even at $r_s$ = 5.
Figure 2 (d) give the g(r) at $r_s$ 5, $T$ =2$T_F$ obtained from
HNC. Unlike in SLTS or
 Tanaka and Ichimaru, the
 $g_{ij}(r)$ is  positive
definite, as expected. Finite-temperature
systems will be discussed more fully in a future publication
as they are relevant to doped-semiconductors and
hot plasmas.\cite{gross}

The $T_q$ determined from the unpolarized $E_c$ 
is used to calculate $E_{xc}(r_s,\zeta, T)$ at arbitrary $\zeta$.
The QMC results for $E_{xc}(r_s,\zeta)$ at T=0
agree with ours, as expected from the
agreement of our $g_{ij}(r)$  with those from MC. 
For example, at $r_s$ = 10, the spin-polarized $E_c$
from Ceperly-Alder is -0.0209 Ry, our HNC procedure gives  -0.0211 Ry,
 Lantto-FHNC gives -0.0186, while
Kallio and Piilo report a value of -0.0171 Ry.\cite{KP}

	In the theory of the UEG the
response $\chi(k,\omega)$ 
of the interacting system is written 
in terms of the noninteracting response
function $\chi^0(k,\omega)$ and a LFC
denoted by $G(k,\omega)$. Considerable effort has been spent in obtaining
even the static approximation $G(k)$ for the LFC. Thus the main
thrust of STLS,\cite{stls}  Ichimaru and Utsumi,\cite{ui} 
Geldart and Taylor,\cite{gt} and others 
has been to provide the $G(k)$
as a function of $r_s$. In the
static case,
for a one-component fluid we have:
\begin{equation}
\label{lfc}
-V_kG(k)=V_k+1/\chi(k)-1/\chi^0(k)
\end{equation}
where $V_k=4\pi/k^2$.
 For a classical fluid we also have
\begin{equation}
S_{ij}(k)=-(1/\beta)\chi_{ij}(k)/(n_in_j)^{1/2}
\label{StoChi}
\end{equation}
Hence the LFCs can be expressed in terms of $S_{ij}(k)$ and compactly
in terms of the short-ranged k-space direct correlation functions 
$\tilde{c}_{ij}(k)$.
For the paramagnetic case:
\begin{equation}
\label{lfcCS}
-V_kG(k)=\{\tilde{c}_{11}(k)+
\tilde{c}_{12}(k)-\tilde{c}^0_{11}(k)\}/2\beta .
\end{equation}
We display the T=0
LFC for $r_s$ =  5 and compare it with the results of QMC and other
theories in Fig.3. In HNC, and in standard SLTS,
the $k$ = 0 limit is overestimated due to the IC. The Vashista-Singwi
version of SLTS, as well as UI have parametrizations which externally
impose the k=0 and k $\to \,\infty$ sum rules. The latter sum rule sates
that $G(k)\to$1-$g(0)$ for large k. The  G(k) from QMC does not
seem to follow this and is close to the 
second-order perturbation calculaton of Geldart and Taylor.\cite{gt}
We have ascertained numerically that our HNC generated G(k)
correctly recovers the k $\to \,\infty$ sum rule
very well.

	In conclusion, we have presented a simple classical mapping
 of a quantum Fermi liquid of arbitrary spin-polarization
and  temperature $T$ and shown that it
{\it quantitatively}$\,$ recovers
 the QMC pair-distribution functions.
The only parameter is a temperature mapping of the
correlation energy.
It seems to be sufficient to map the static
properties of the UEG. Using the method, we have examined
spin-dependent correlation energies, pair-correlation functions at zero and
finite T, as well
as the local-field correction to the response functions. The method
clearly has
potential applications to Bose fluids, 2-D electron and other systems.


\newpage

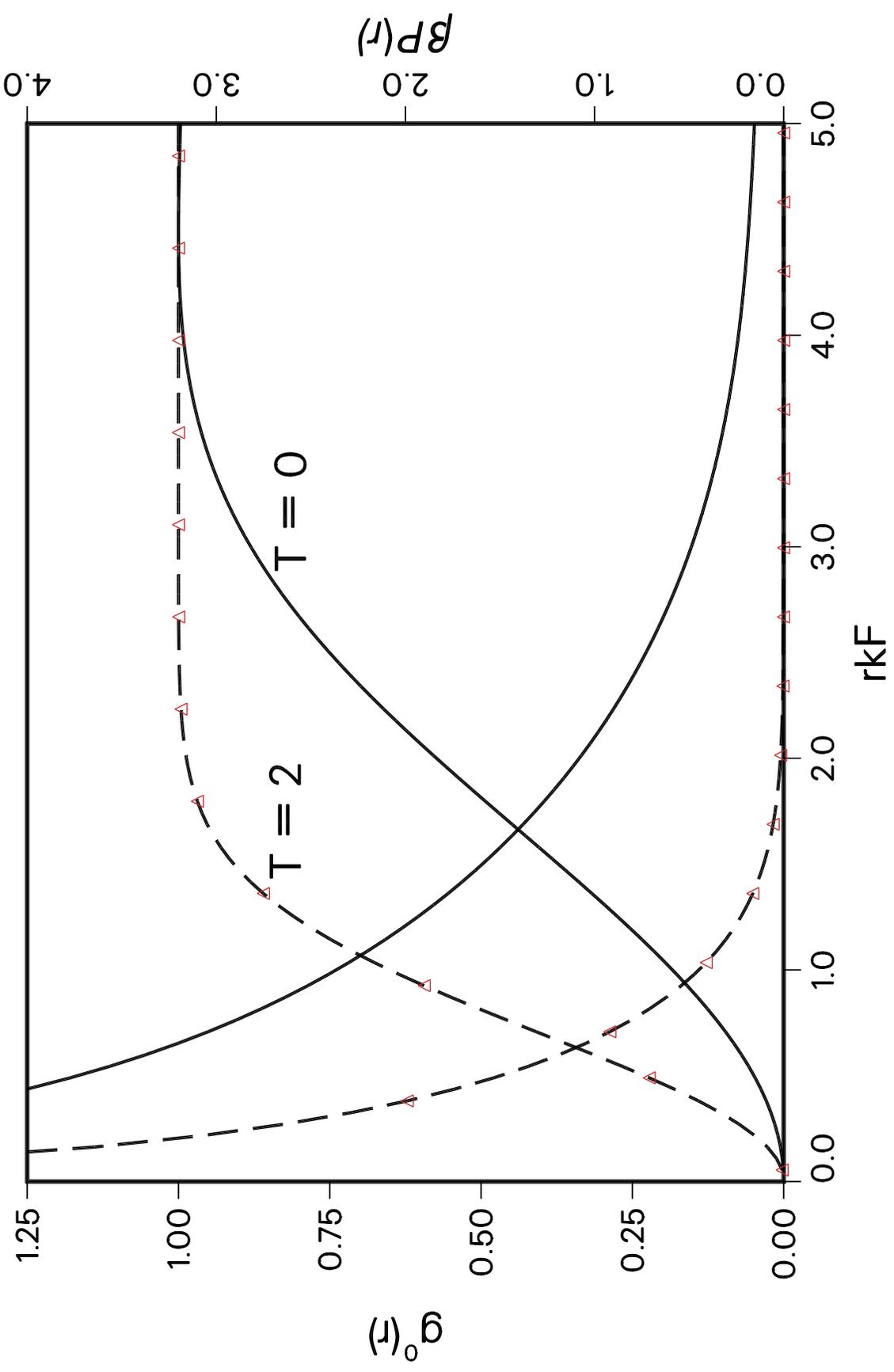
\begin{figure}
\caption
{The non-interacting PDF, $g^0_{11}(r)$, and the
Pauli potential, Eq.~8, at $T/E_F$=0 and 2 (dashed lines).
They are universal functions
of $rk_F$, where $k_F$ is the Fermi wavevector.
}
\label{fig1}
\end{figure}

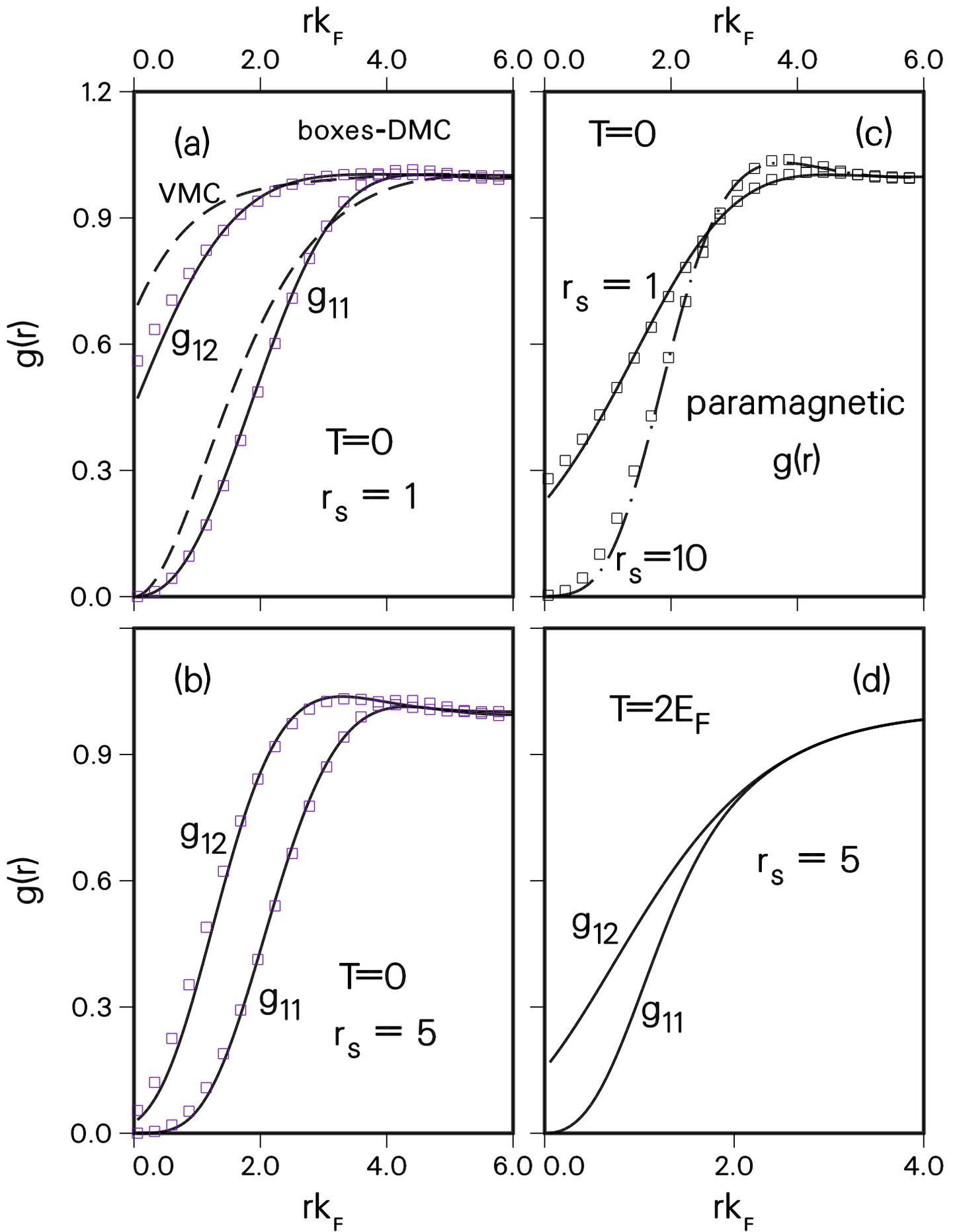
\begin{figure}
\caption
{The interacting PDFs $g_{11}(r)$ and 
$g_{12}(r)$ at $r_s$=1 . Solid lines-HNC, boxes:-DMC (a),
 HNC, VMC (dashes),and DMC.  Panel (b), $r_s=5$, DMC
and HNC (c), (d) are for $T/E_F$=2. In (c) The paramagnetic
$g(r)$ at $r_s$=1 and 10 at T=0 compared with the DMC simulations.
}
\label{fig2}
\end{figure}

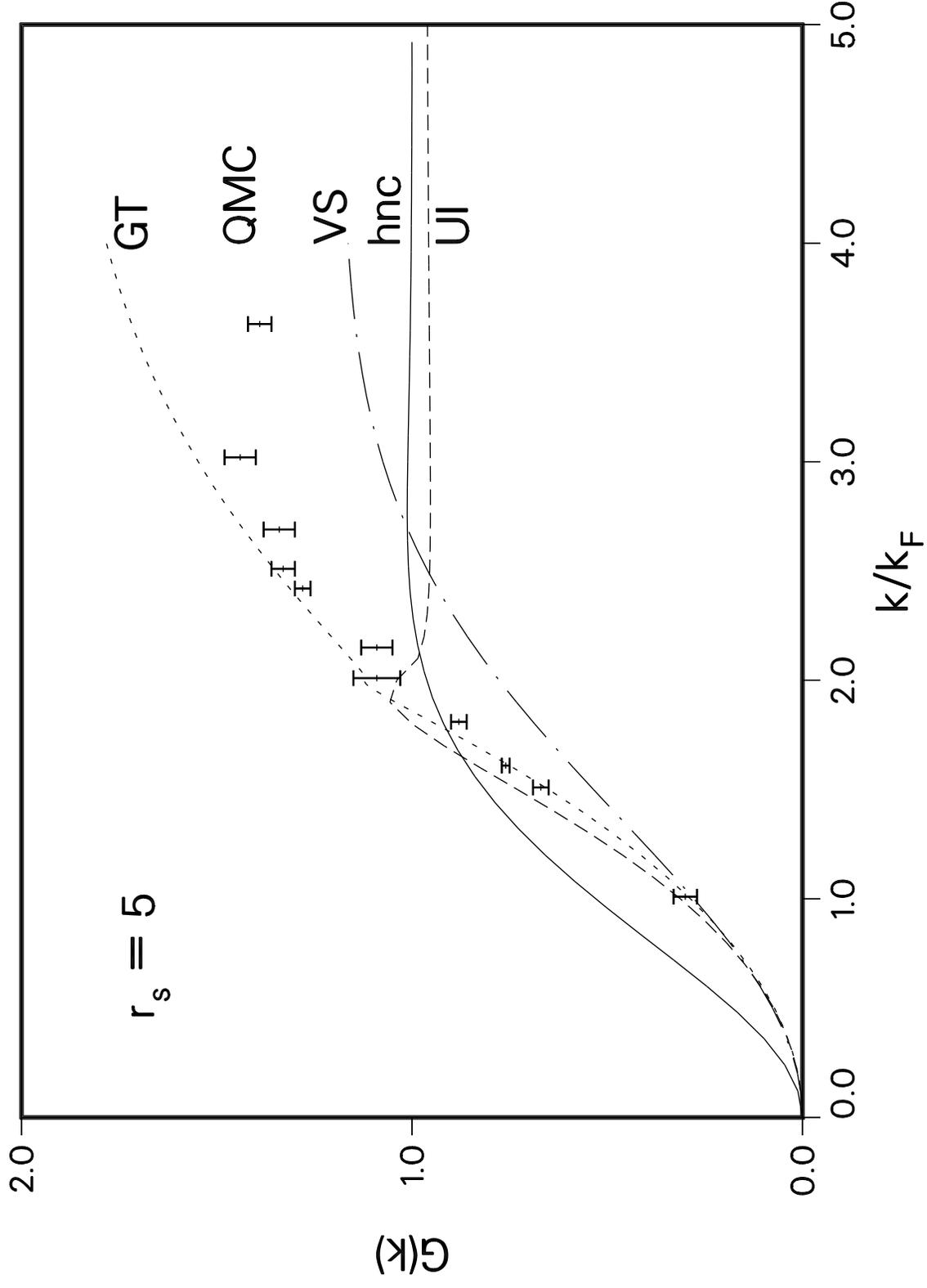
\begin{figure}
\caption
{The local-field correction $G(k)$, Eq.~\ref{lfcCS},
to the static rsponse at T=0 and $r_s$ = 5.
Results from the HNC, QMC,\cite{mcs}
Geldart and Taylor (GT),\cite{gt}
Vashista and Singwi(VS),\cite{stls}, and Utsumi and Ichimaru (UI),\cite{ui}
are shown.
}
\label{fig3}
\end{figure}


\begin{references}
\bibitem[\dag]{byline1} electronic mail address: chandre@cm1.phy.nrc.ca
%
\bibitem{textbk}
A. Fetter and J. Walecka {Quantum theory of Many-Particle systems}
(McGraw-Hill, New York) 1971
G. D. Mahan,{\em Many-Particle Physics} (Plenum, New York) 1981
%
\bibitem{K-S}
N.D. Mermin, Phys. Rev.{\bf 137}, A1441 (1965); P. Hohenberg and W. Kohn, Phys.
Rev. {\bf 136},B864 (1964);
 W. Kohn and L.J. Sham, Phys. Rev. {\bf 140}, A1133 (1965).
%
\bibitem{gt}
W. Geldart and R. Taylor, Can. J. Phys. {\bf 48},155 (1970);
 and
references therein.
\bibitem{ahm}
A. H. MacDonald, M. W. C. Dharma-wardana and W. Geldart, J. Phys. F
{\bf 10} 1919 (1980)
\bibitem{green}
F. Green, D. Neilson and J. Szyma\'nski, Phys. Rev. B {\bf 31} 2796 (1985);
Tanaka and Ichimaru, Phys. rev. B {\bf 39}, 1036 (1989);
Richardson and Ashcroft, Phys. Rev. B (1995)
%
%
\bibitem{toigo}
Toigo and Woodroff, Phys. Rev B, {\bf 4} 371 (1971);
M. W. C. Dharma-wardana and R. Taylor, J. Phys. F {\bf 10} 2217 (1980)
%
\bibitem{stls}
K. S. Singwi, M. P. Tosi, R. H. Land, and A. Sj\"olander, Phys. Rev. {\bf 176}
589 (1968);
P. Vashista and K. S. Singwi, Phy. Rev. {\bf 6} 875 (1972)
%
\bibitem{italians}
S. V. Adamjan, J. Ortner, and I. M. Tkachenko, Europhysics lett. {\bf 25}
11 (1994); V. Contini, G. Mazzone and F. Sacchetti, Phys. Rev. B {\bf 33},
712 (1986)

\bibitem{ui}
K. Utsumi and S. Ichimaru, {\bf 24} 7385 (1981)


\bibitem{feenberg}
E.Feenberg, {\em Theory of Quantum Fluids} (Academic, New York 1969);
C. E. Campbell and J. G. Zabolitsky, Phys. Rev. B {\bf 27}, 7772 (1983);
T. Chakraborty, Phys. Rev. B {\bf 29}, 1 (1984)
%
\bibitem{lantto}
L. J. Lantto,  Phys. Rev. B {\ bf 22}, 1380 (1980) 
%
\bibitem{KP}
A. Kallio and J. Piilo, Phys. Rev. Lett. {\bf 77}, 4237 (1996)
%
\bibitem{ceperley81}
D. M. Ceperley, {\em Recent Progress in Many-body Theories}, Ed. J. B.
Zabolitsky (springer, Berlin 1981), p262;
\bibitem{ortiz}
G. Ortiz abd P. Ballone, Phys. Rev. B {\bf 50}, 1391 (1994)
%
\bibitem{mcs} by 
S. Moroni, D. Ceperley and G. Senatore, Phys. Rev. Lett, {\bf 75} 689 (1995),
also C. Bowen, G. Sugiyama and B. J. Alder, Phys. Rev. {\bf 50}, 14838 (1994)
%
%
\bibitem{vwn}
S. H. Vosko, L. Wilk and M. Nusair,  Can. J. Phys.{\bf 25},283 (1989),
%
\bibitem{p-z}
J. P. Perdew and A. Zunger, Phys. Rev. B {\bf 23},5048 (1981) 
%
\bibitem{hncref}
J. M. J. van Leeuwen, J. Gr\"oneveld, J. de Boer, Physica {\bf 25}, 792 (1959)
and T. Morita, and H. Hiroike, Prog. Theor. Phys. {\bf 23}, 1003 (1960)
%
\bibitem{lado}
 F. Lado, J. Chem. Phys. {\bf 47}, 5369 (1967)
%
\bibitem{rosen}
F. Lado, S. M. Foiles and N. W. Ashcroft, Phys. Rev. {\bf A 26}, 2374 (1983)
Y. Rosenfeld, Phys. Rev. A {\bf 35}, 938 (1987)
%
\bibitem{kittel}
C. Kittel, {\it Quantum theory of Solids} p 94, (Wiley, New York) 1987
%
%
%
%
\bibitem{minoo}
M. Minoo, M. Gombert and C. Deutsch, Phys. Rev. A {\bf 23}, 924 (1981)
%
%
\bibitem{p-w}
J. P. Perdew and Y. Wang, Phys. Rev. B {\bf 45} 13244 (1992)
%
\bibitem{gross}
M.W.C. Dharma-wardana and F. Perrot p 635 in {\em Density Functional
Theory}, Edited by E. K. U. Gross and R. M. Dreizler (Plenum, New York) 1995
%
%

%





\end{references}
\end{document}